\documentclass[onecolumn]{article_arxiv}

\usepackage[utf8]{inputenc}
\usepackage{mathtools}
\usepackage{hyperref}
\usepackage{graphicx}
\usepackage{dcolumn}
\usepackage{bm}
\usepackage{epsf}
\usepackage{subfigure}
\usepackage{epstopdf}
\usepackage{amsmath}
\usepackage{amssymb}
\usepackage{color}

\newcommand{\nth}{$n^{\textrm{th}}$ }
\usepackage{braket}

\newcommand{\kb}{k}
\newcommand{\kt}{\kb T}
\newcommand{\vm}{\sigma_{\rm m}^2}
\newcommand{\ud}{\mathrm{d}}

\newcommand{\p}{{\mathcal P}}

\title{Reversible feedback confinement} 

\author{L\'eo Granger\inst{1} \and Luis Dinis\inst{1}\and Jordan M. Horowitz\inst{2}\and Juan MR Parrondo\inst{1}}

\date{\today}

\begin{document}

\maketitle

\institute{
    \inst{1} Departamento de F\'isica At\'omica, Molecular y Nuclear
    and  GISC, Universidad Complutense Madrid, 28040 Madrid, Spain\\
    \inst{2} Department of Physics, Physics of Living Systems,
Massachusetts Institute of Technology, Cambridge, MA 02139, USA
}
\abstract{
We present a feedback protocol that  is able to confine a system to a
single micro-state without heat dissipation. The protocol  adjusts
the Hamiltonian of the system in such a way that the Bayesian
posterior distribution after measurement is in equilibrium. \revision{As a result, the whole process satisfies feedback reversibility -- the process is indistinguishable from its time reversal -- and assures the lowest possible dissipation for confinement. In spite of the 
whole process being reversible it can  surprisingly be implemented in
finite time}. We illustrate the idea with a Brownian particle in a
harmonic trap with increasing stiffness and present a general theory
of reversible feedback confinement for systems with discrete states.
}

\fintitulo

Micro-manipulation  techniques introduced in the last decades ---optical and
magnetic tweezers, monitored quantum dots, or atomic force microscopy--- call
for a theoretical framework to analyze driven systems at the micro-scale. An
important element in such a framework is feedback control, where the system is
monitored or measured at certain stages of a process, and the driving protocol
depends on the outcomes of the measurements \cite{Touchette2000}. Feedback control can be used, for
instance, to increase the performance of certain devices, such as Brownian
ratchets \cite{cao_feedback_2004,Lopez:2008gl} and micro-motors
\cite{abreu_extracting_2011}, or to diminish thermal fluctuations, confining
the system to a small region of phase space. A particular case of the latter is
feedback cooling, a  technique that has been implemented  in classical and
quantum systems \cite{Li:2011jl,Gieseler:2012bi,Cohen2005,Cohen2006} and plays
a relevant role in quantum optomechanics \cite{Aspelmeyer:2014ce}.

Now, it is well established that for any feedback protocol the information
itself is a thermodynamic resource, which is quantified within the framework of
information thermodynamics~\cite{Parrondo:2015cv}. In particular, the work $W$
needed to perform an isothermal feedback  process at temperature $T$ is bounded
by
\cite{sagawa_second_2008,cao_thermodynamics_2009, allahverdyan_thermodynamic_2009,horowitz_nonequilibrium_2010,fujitani_jarzynski_2010,ponmurugan_generalized_2010,ito_information_2013,shiraishi_fluctuation_2015,horowitz_thermodynamics_2014,hartich_stochastic_2014,granger_irreversibility_2013,Parrondo:2015cv}
\begin{equation}
	W \ge \Delta F - \kb T I,
	\label{e.2ndlawfb}
\end{equation}
where $\kb$ is Boltzmann's constant, $\Delta F$ is the
difference in free energy between the final and initial
states of the system, and $I$ is the amount of information 
gained by the measurement. For a single feedback loop, $I$ is the information-theoretic {\em
mutual information} between the measurement outcome and the
system's micro-state~\cite{Parrondo:2015cv,sagawa_second_2008}.
As a consequence of (\ref{e.2ndlawfb}), the work $W$ performed in a
feedback process might be lower than the variation in the
free energy of the system $\Delta F$.

Processes saturating bound (\ref{e.2ndlawfb}) are of particular relevance as
they make an optimal use of  information. For a single measurement, such
processes were identified and characterized in
\cite{horowitz_thermodynamic_2011,horowitz_designing_2011,horowitz_optimizing_2013},
\revision{where it is shown that optimal processes must be reversible under feedback}. For
systems in contact with a thermal reservoir, reversibility requires that the
probabilistic state of the system conditioned on the measurement outcome be the
equilibrium Gibbs state at all times. In particular, after a measurement, the
probabilistic state changes abruptly: it is updated via Bayesian inference
using the information provided by the measurement \cite{Parrondo:2015cv}. To
achieve reversibility, the Hamiltonian of the system must be tuned so that its
Gibbs distribution coincides with the  measurement-dependent posterior
probability distribution~\cite{esposito_second_2011,jacobs_quantum_2012}.

In this letter, we apply this recipe to design an optimal, reversible feedback
protocol that confines a system in contact with a thermal bath to a small
region of phase space or even to a single micro-state.  In any such confinement
procedure, the free energy change is $\Delta F=\Delta E-T\Delta S$ where
$\Delta E$ is the change of internal energy of the system and $\Delta S$ the
change in the entropy, which is negative. Hence, \eqref{e.2ndlawfb} can be
rewritten as
\begin{equation}\label{e.2ndconf}
Q_{\rm diss} =W-\Delta E\geq -T\Delta S-\kb  TI.
\end{equation}
Without feedback ($I=0$), confinement must be accompanied by dissipated heat
$Q_{\rm diss}\geq -T\Delta S>0$, which is a direct consequence of the second
law~\cite{esposito_second_2011,Horowitz2013c}.  This dissipation, however, can
be decreased if we make use of  feedback. 
Here we introduce an optimal feedback confining protocol
consisting of a series of measurements, each followed by a rapid tuning of the
Hamiltonian that leaves the system in equilibrium. Between two consecutive
measurements the Hamiltonian is kept constant.  Reversibility ensures that equality is met in
 \eqref{e.2ndconf}. In addition, since the
probabilistic state of the system does not change between measurements, the
amount of heat dissipated to the environment is zero. Hence, we are able to
design a general protocol where the reduction in the system's entropy is
entirely due to the acquisition of information, $-\Delta S=kI$.

We first illustrate this idea with a simple example, similar to the one
introduced in \cite{abreu_extracting_2011}.  Consider an overdamped Brownian
particle with position $z(t)$ in a harmonic potential $V_{\kappa,y}(z) =
\frac{1}{2}\kappa\left(z-y  \right)^2$ with tunable center $y$ and stiffness
$\kappa$.  The particle's dynamics obey the Langevin equation $\gamma \dot z= -
\kappa ( z- y) + \xi_t$, where $\gamma$ is the friction coefficient and $\xi_t$
is zero-mean Gaussian white noise with covariance $\langle \xi_t \xi_{t'}
\rangle = 2\gamma \kb T \delta(t - t')$ verifying the fluctuation-dissipation
theorem.
\begin{figure}
	\includegraphics[scale=.51]{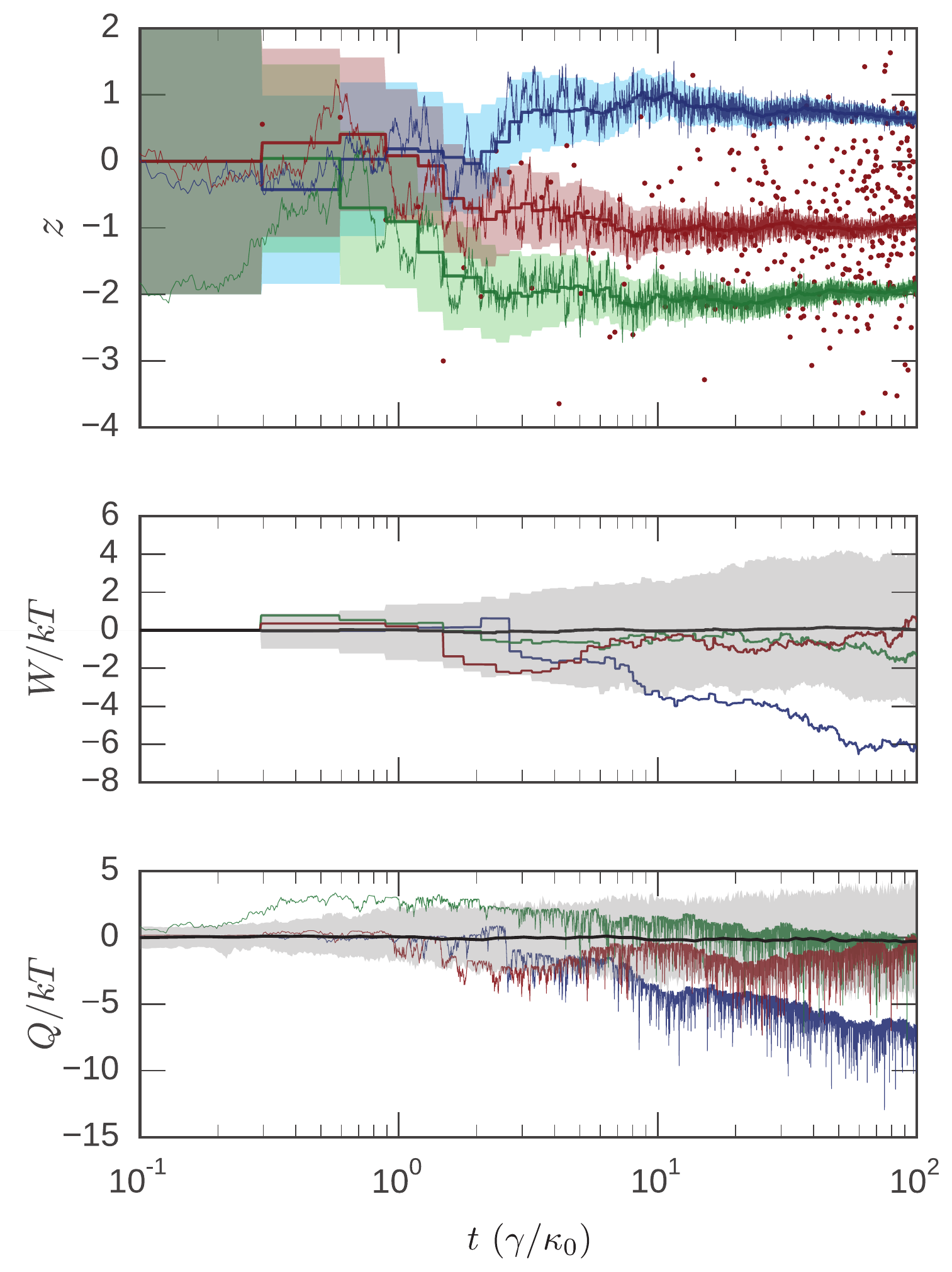}
	\caption{(Color online) {\em Top panel:} Three realizations (blue, red,
	green) of reversible confinement for a harmonically-trapped Brownian
	particle.  The strongly-fluctuating thin lines represent the particle's
	time-dependent position $z(t)$; the thick lines are the trap centers
	$y_t$; and the shaded area corresponds to $y_t \pm 2 \sigma_t$, where
	$\sigma_t = \kt/\kappa_t$ is the equilibrium variance. The scatter of
	red dots are the red trajectory's measurement outcomes.  Measurements
	are performed every $\Delta t = 0.3(\gamma/\kappa_0)$ with error
	$\vm=\sigma_0^2=1$.  {\em Middle panel:} Work performed during the
	confinement process. The black line is the average over 100
	realizations and the gray shaded area is the middle 90\% of the work
	values. The colored lines correspond to the realizations depicted in
	the top panel. {\em Bottom panel:} Same as middle for the heat released
into the environment.}
	\label{f.ex}
\end{figure}

In the absence of control, the particle's position relaxes to equilibrium with
Gibbs distribution,
\begin{equation}
	\rho_{\kappa,y}(z) \propto \exp\left(
	-\frac{V_{\kappa,y}(z)}{\kb T}
	\right),
	\label{e.bp.eq}
\end{equation}
which, in this case, is Gaussian with mean $\bar z = y$ and variance $\sigma^2
= \kb T / \kappa$.  Our goal is then to confine the particle to a smaller
region of space by manipulating the control parameters $\lambda=(\kappa,y)$
using feedback.

To implement the feedback, we preform an imperfect measurement of the particle's position, yielding outcome $m$ with distribution $q(m|z)$ depending on the particle's true position $z$.
Here, we make the standard assumption~\cite{abreu_extracting_2011} of Gaussian errors, where $m$ is drawn from the Gaussian distribution about $z$,
\begin{equation}
	q(m|z) \propto \exp\left(-\frac{\left( m - z \right)^2}{2\vm}\right),
	\label{e.gausserr}
\end{equation}
where the standard deviation $\sigma_{\rm m}$ quantifies the error in the measurement.

The particle's post-measurement distribution $\rho'$  is
obtained from the pre-measurement distribution $\rho$ via Bayes' theorem:
\begin{equation}
	\rho'(z|m) = \frac{q(m|z)\rho(z)}{\pi(m)},
	\label{e.bayesBP}
\end{equation}
where $\pi(m) = \int q(m|z)\rho(z)\ud z$ is the marginal distribution of
measurement outcomes.  For the control to be reversible, we need to adjust our
control parameters $\lambda$ immediately after the measurement, so that
$\rho^\prime$ is the new equilibrium state.
It is straightforward to check that, after a measurement characterized by the Gaussian 
distribution \eqref{e.gausserr}, a Gaussian pre-measurement state, like
\eqref{e.bp.eq}, centered at $\bar z$ with variance $\sigma^2$, updates to a
post-measurement state according to \eqref{e.bayesBP} that is also Gaussian
with mean and variance \cite{abreu_extracting_2011}
\begin{eqnarray}
	\bar z'(m) &=&  \frac{\vm}{\sigma^2+\vm} \bar z +
	\frac{\sigma^2}{\sigma^2+\vm} m 
	\label{e.postMean}\\
	\frac{1}{\sigma'^2} &=&  \frac{1}{\sigma^2} + \frac{1}{\vm}.
	\label{e.postVar}
\end{eqnarray}
Thus, we can make the post-measurement distribution an equilibrium
distribution by setting the stiffness and the center of the trap to the new
values $\kappa' = \kt / \sigma'^2$ and $y'(m) = \bar z'(m)$.

The effect of repeating this reversible feedback procedure at successive times
$t_1,\dots,t_n$ is to confine the particle at a well-defined position.  In
fact, (\ref{e.postVar}) implies that the post-measurement stiffness $\kappa' =
\kappa + \kt / \vm > \kappa$ increases by $kT/\vm>0$ in each reversible
feedback step.  Therefore, the post-measurement stiffness $\kappa_n$ after the
$n^{\textrm{th}}$ measurement at time $t_n$ is 
\begin{equation}
	\kappa_n = \kappa_0 + n\frac{\kt}{\vm},
	\label{e.kappan}
\end{equation}
where $\kappa_0$ is the initial stiffness of the trap.  Consequently, the
post-measurement variance  of the particle $\sigma_n^2 = \kt / \kappa_n$
decreases to zero as the number of measurements increases.

All along the process the average potential energy of the particle is constant
and equal to $\frac{1}{2}\kb T$, by the equipartition theorem\revision{,} so $\Delta E=0$. 
\revision{In addition, as discussed above, since the potential is instantaneously modified to leave the particle at equilibrium, there is no subsequent evolution nor relaxation between measurements and $Q=0$. Hence, the first law implies $W=0$ on average.

Alternatively, we can confirm there is zero heat dissipation and that the process is reversible directly from the second law by  computing the mutual information between $m$ and $z$ in one measurement,} 
\begin{equation}
  \revision { I(m,z)=\int\!dm dz\,\pi(m)q(m|z)\log{\frac{q(m|z)}{\pi(m)}}},
  \label{e.mutual_info_G}
\end{equation}
\revision{and demonstrating its equality with $-\Delta S$.}

\revision{In our example, $q(m|z)$, $\rho(z)$ and $\pi(m)$ are Gaussian distributions of variances $\sigma_m^2$, $\sigma^2$ and $\sigma_m^2\sigma^2/\sigma'^2$. Using the entropy of a Gaussian distribution $\rho(X)$, $h(\rho)=\log(\sigma_x\sqrt{2\pi e
})$,  it is straight forward to show that
  \begin{equation}
    I(m,z)=\frac{1}{2}\log{\frac{\sigma'^2}{\sigma^2}.}
  \end{equation}
  On the other hand, after a measurement, the state of the system updates from a Gaussian of variance $\sigma^2$ to a Gaussian of variance $\sigma'^2$ as shown in \eqref{e.postVar}. Hence, the entropy of the particle changes as
  \begin{equation}
    \begin{split}
    \Delta  S &=-k \left(\log(\sigma'\sqrt{2\pi e})-\log(\sigma\sqrt{2\pi e}\right)=-k\frac{1}{2}\log{\frac{\sigma'^2}{\sigma^2}}\\ &=-kI(m,z).
  \end{split}
  \end{equation}
  Since the particle is always at equilibrium, the second law \eqref{e.2ndconf} applies with an equal sign and $Q=0$.

The optimal protocol thus ensures that} no work
is performed nor heat is dissipated on average. On the other hand, both would
diverge for any non-feedback protocol that squeezes the trap potential to a
single point, since in this case  $\Delta S$ is infinity.  Another important
feature is that the confinement can be realized as fast as we like: the
potential is updated immediately after each measurement and the system's
probabilistic state, being always in equilibrium, does not change between
measurements. Hence, there is no need of running the protocol slowly to achieve
reversibility.

Three realizations of the trapping process with imprecise measurements
($\vm=\sigma_0^2=1$) are depicted on Fig.~\ref{f.ex}.  The top panel shows the
time evolution of the position of the particle (thin strongly-fluctuating
lines) and of the trap (thick lines). For the realization ending in the middle
(red), we also show the measurement outcomes.  Even though the measurement is
not very precise, confinement is achieved eventually. In fact, each measurement
provides a piece of information, even if it might be small. The middle and the
bottom panels respectively depict the work performed on the particle and the
heat released to the environment. On these panels, the black line represents
the average over 100 realizations and the gray shaded area encompasses the
middle 90\% of work and heat values (notice that the blue realization is
atypical from an energetic point of view).  We indeed verify that the average
heat and work are zero during the process.

We now turn to the formulation of a general theory of reversible
feedback confinement for a system with discrete phase space $\Gamma$ in contact
with a heat bath at temperature $T$. The
generalization to continuous phase space is straightforward.
Assume that the energy $E(z|\lambda)$
of any micro-state $z \in \Gamma$ can be controlled through the
parameter (or set of parameters) $\lambda$.
As in our example, the system is initially in equilibrium, with control parameter set to an initial value $\lambda = \lambda_0$.  
Then, at times $t_1,\dots,t_N$ we measure a
system observable $M=M(z)$ with outcomes drawn from the set $m\in{\mathcal M}$ according to the distribution $q(m|z)$, followed by a measurement-dependent switch of $\lambda$ to perform a reversible feedback process.

After the \nth measurement, with outcome $m_n$, the post-measurement distribution $p_n$ obtained from the pre-measurement state $p_{n-1}$ according to Bayes' theorem is
\begin{equation}
	p_n(z|\mu_n) =
	\frac{q(m_n|z)p_{n-1}(z|\mu_{n-1})}{\pi_n(m_n|\mu_{n-1})},
	\label{e.posterior}
\end{equation}
where $\mu_n \equiv (m_1, \dots, m_{n-1},m_n)$ is the list of the first $n$ 
measurement outcomes and $\pi_n(m_n|\mu_{n-1}) = \sum_z
q(m_n|z)p_{n-1}(z|\mu_{n-1})$ is
the probability that the outcome of the \nth measurement is $m_n$ given 
the list $\mu_{n-1}$ of the first $n-1$ outcomes. 
Notice that here we explicitly indicate the dependence 
of the posterior distribution $p_n(z|\mu_n)$ on the whole history of measurements $\mu_n$ [cf.~\eqref{e.bayesBP}].

The reversible feedback condition dictates that immediately after the
measurement the control parameter is changed to a value $\lambda(\mu_n)$  such that the post-measurement distribution
(\ref{e.posterior}) is equilibrium:
\begin{equation}
	E(z|\lambda(\mu_n)) = -\kt \log p_n(z|\mu_n) + F_n(\mu_n),
	\label{e.revfb}
\end{equation}
where $F_n(\mu_n)$ is the free energy of the system after the switch.
Its value is arbitrary and can be changed by redefining the reference energy, affecting only the work and not the dissipated heat.
Between two measurements, the control parameter is held constant.
Since the system is in equilibrium, the distribution $p_n$ is invariant between
times $t_n^+$ and $t_{n+1}^-$. 
Therefore it is also the pre-measurement distribution of the subsequent
$(n+1)^{\textrm{th}}$ measurement.

The effect of the procedure described above is to eventually confine the system
into one single micro-state (or set of micro-states) without dissipating heat to the environment.
This confinement relies on the following remarkable property of the process\revision{, which we will demonstrate in the following (cf.~\eqref{e.rec} to \eqref{e.prod})}:
given the final micro-state $z_N=z(t_N)$, the measurement outcomes in the sequence $\mu_N =
(m_1,\dots,m_N)$ are \emph{independent and identically
dnistributed} (i.i.d.) according to $q(m|z_N)$. 
Thus, by the law of large numbers, when the number of measurements $N$ is large, the empirical frequency $\nu_m$ of outcome $m$ ---that is the fraction of times $m$ appears in the list $\mu_N$--- is close to $q(m|z_N)$.
As a consequence, just knowing the empirical distribution $\nu=\{\nu_m\}$ of the list $\mu_N$ of measurement outcomes, we can infer with certainty that the final micro-state $z_N$ is in the set of micro-states that verify $q(m|z_N) \simeq \nu_m$ for all $m$.
If the distributions $q(m|z)$ are different for all micro-states $z$, then the system is confined to a single micro-state.
On the other hand, if there exist two micro-states $z^\prime$ and $z^{\prime\prime}$ such that $q(m|z^{\prime})=q(m|z^{\prime\prime})$, then the two states are indistinguishable to the observer, and our protocol confines the system to the set $\{z^{\prime},z^{\prime\prime}\}$.
To arrive at  these conclusions, we now calculate the distribution $p_N(z|\nu)$ of $z_N$ conditioned on the empirical density $\nu$ and show that it is highly peaked.

To begin, we
apply (\ref{e.posterior}) recursively:
\begin{equation}
	p_N(z|\mu_N) = \frac{p_0(z) \prod_{n=1}^N
q(m_n|z)}{\pi_N(m_N|\mu_{N-1})\dots \pi_1(m_1)},
	\label{e.rec}
\end{equation}
where $p_0(z)$ is the initial equilibrium distribution.
The denominator \eqref{e.rec} is simply the probability to observe the sequence $\mu_N$. 
Therefore, the joint probability of observing $\mu_N$ and ending the process at $z_N = z$ reads
\begin{equation}
	\p_N(z,\mu_N) = p_0(z) \prod_{n=1}^N q(m_n|z).
	\label{e.joingen}
\end{equation}
Summing over all possible $\mu_N$, we obtain the marginal distribution of $z_N$, $p_N(z)=p_0(z)$, which remarkably is  constant along the whole process. 
Thus, the probability of $\mu_N$, given the final micro-state $z_N$ is
\begin{equation}
	P_N(\mu_N|z_N) = \prod_{n = 1}^N q(m_n|z_N) = \prod_{m\in{\mathcal M}}
	q(m|z_N)^{N\nu_m}.
	\label{e.prod}
\end{equation} 
That is, given the final state $z_N$, the measurement outcomes are i.i.d., so that the probability to observe a specific sequence $\mu_N$ depends only on the empirical distribution $\nu=\{\nu_m\}$.
The probability to observe an empirical distribution $\nu$, given $z_N$, 
is obtained  by multiplying  \eqref{e.prod} by the number of sequences yielding the same distribution,
\begin{equation}
	P_N(\nu|z_N) = \frac{N!}{\prod_m (N \nu_m)!}P_N(\mu_N|z_N)
	\approx e^{ -N \epsilon(\nu,z_N) }
\end{equation}
for $N$ large, where we used Stirling's approximation and defined
$\epsilon(\nu,z)=D[\nu_m\|q(m|z) ]= \sum_m \nu_m \log \frac{\nu_m}{q(m|z)}$,
the relative entropy between the empirical distribution $\nu$ and
$q(m|z)$~\cite{Cover}. 
As a relative entropy, $\epsilon=0$ only when $\nu_m=q(m|z)$.
Notice that $\epsilon(\nu,z_N)$ is  the large deviation
rate function for the empirical distribution $\nu$ \cite{Dembo}.

Finally, applying again Bayes' theorem, we get the probability that the final
state is $z_N=z$ given the empirical distribution $\nu$ of measurement
outcomes,
\begin{equation}
	p_N(z|\nu) = \frac{ P_N(\nu|z) p_0(z)}{\sum_{z'}
P_N(\nu|z')p_0(z')}
\approx \frac{p_0(z) e^{-N \epsilon(\nu,z)}}{\sum_{z'}p_0(z') e^{-N \epsilon(\nu,z')}}.
\end{equation}
For this distribution to be equilibrium, the final energy landscape must be
\begin{equation}
E(z|\lambda(\mu_N))=E(z|\lambda_0)+NkT\epsilon(\nu,z).
\end{equation}
For large $N$, the system becomes trapped in the most likely micro-state(s)
$z^*$ that minimize $\epsilon(\nu,z^*)$.  Those are exactly the states for
which the empirical frequencies $\nu_m$ approach $q(m|z^*)$, making
$\epsilon(\nu,z^*)\to 0$.  As in our example, the probabilistic state of the
system is constant between measurements and therefore the average dissipated
heat is zero.

We have shown that it is possible to use feedback to confine a system with zero
average heat, even when the measurements have large errors.  In contrast, to
achieve the same confinement without feedback, one has to dissipate an enormous
amount of heat. 

One of the main novelties of the present work is the consideration of errors in
the measurement.  With error-free measurements, confinement without dissipation
is trivial. However, we have shown that confinement is still possible no matter how
inaccurate the measurements are; although for less precise measurements more are needed to achieve perfect confinement. \revision{For instance, in our Brownian particle example, the variance of the particle after $n$ measurements can be obtained by recurring \eqref{e.postVar} giving $(\sigma^n)^2=\sigma_m^2\sigma_0^2/(n\sigma_0^2+\sigma_m^2)$, with $\sigma_0^2$ the initial variance of the particle. For large $n$, $(\sigma^n)^2\approx\sigma_m^2/n$. Thus, to confine the particle, that is, to reduce particle variance below a small target value $(\sigma^n)^2<\delta^2$, the number of measurements needed are proportional to the measurement error $\sigma_m^2$: $n\gtrsim \sigma_m^2/\delta^2$. }

Errors in the
measurements can also give rise to counter-intuitive results, such as providing better discrimination of the system's micro-states.
Consider, for instance, a particle in a one-dimensional box of length 1 and  a
binary measurement with $m\in\{\textrm{L, R}\}$, standing for the left and right half of the box,
respectively. Our protocol with an error-free measurement, $q(\textrm{R}|z)=1$
if $z>1/2$ and zero otherwise and  $q(\textrm{L}|z)=1-q(\textrm{R}|z)$, will
confine the particle to one of the two halves of the container and will not be
able to discriminate micro-states within each half of the container. \revision{Notice that in this case, although measurement outcome $\{L,R\}$ depends on the state $z$, there are a number of different states that share a common outcome probability distribution $q(m|z)$ and, as stated above, the observer cannot distinguish among them.} On the
other hand, a binary measurement with $q(\textrm{R}|z)=z$ and
$q(\textrm{L}|z)=1-z$ is able to confine the particle to a single point with
zero heat.

We conclude with some remarks.  
First, while the system dissipates no heat, there is an energetic cost
to acquiring or, alternatively, to erasing the information
\cite{Parrondo:2015cv,sagawa_second_2008,sagawa_minimal_2009,granger_thermodynamic_2011,granger_differential_2013}.
There is no
avoiding the second law, we simply have shifted its cost to the memory device.
However, such a shift is relevant when we need to keep low the  heat dissipated to the surroundings of the controlled system\cite{kutvonen_thermodynamics_2016,koski_-chip_2015}. 

Second,
although the final state of any single realization is tightly confined, that
final state is {\em a priori} random: the marginal distribution of the system
at the end of the confinement process is the same as the initial distribution.
The trade-off is that our protocol can be implemented at arbitrary speed
without compromising  reversibility.
The reason is that the statistical state
of the system is frozen between measurements, so that the system
\emph{never actually relaxes towards equilibrium}. Hence, its relaxation time
is irrelevant and 
the speed of the process
is simply given by the frequency of the measurements. However, if we wished to confine
the system to a fixed pre-determined low-entropy state, as required for
feedback cooling~\cite{Touchette2000, Horowitz2013c}, our protocol requires one last slow isothermal step that
drags the system to the target state.   To be precise, even though the relaxation time to equilibrium can be avoided using our confinement protocol, the speed of the process could be still limited by other factors, for instance, the time needed to modify the Hamiltonian experimentally. In many systems though, like a colloidal particle trapped in a laser beam for example, the tuning of the Hamiltonian can actually be done much faster than the position relaxation time of the particle. Our protocol would then ensure a very low dissipation at fast operation times.

Lastly, our protocol requires fine-tuned
control over the Hamiltonian to fulfill~\eqref{e.revfb}.  This can be a hinderance
if we aim to decrease the fluctuations of the velocity of a classical particle~\cite{Kim2007,Munakata2012},
since it is difficult to construct  velocity-dependent forces in the lab.  This
difficulty could be overcome in the quantum version of our procedure, since
fine-tuned control over the energy-levels of small quantum systems  is often
more feasible. For instance, increasing the stiffness of a harmonic trap in the
quantum regime, affects both the kinetic and the potential energy of the
oscillator. 
Regarding the fine-tuning of the Hamiltonian needed for confinement, our example provides also an interesting lesson. As indicated by equation \eqref{e.kappan}, there is a trade-off between the measuring error and the fractional change of the control parameter needed to adjust the Hamiltonian at each step. Therefore, when it is experimentally difficult to slightly modify the control parameter, our result suggest this could be overcome also by increasing measuring precision.

Interestingly, our protocol being reversible can also be employed to extract the maximum amount of work from the information provided by a number of measurements\revision{, it suffices to perform an isothermal quasistatic (reversible) transformation from the confined state back to the initial unconfined state, i.e., a sort of {\em expansion}. During reversible confinement without work, the system increases its free energy in an amount $\Delta F=kTI$. By expanding reversibly and isothermally (without any further measurement) all this free-energy is converted into (extracted) work $W=-\Delta F=-kTI$. In the first part of the operation, the optimal feedback transforms the information contained in fluctuations into free energy but, by confining the system, reduces the fluctuations exhausting the amount of information. Then, during work extraction, heat is allowed to enter the system which in turn develops new fluctuations that can be exploited in the next cycle of operation \cite{Bauer2012,abreu_extracting_2011,horowitz_imitating_2013,sandberg_maximum_2014,horowitz_second-law-like_2014}. }


This work has been supported by grants ENFASIS (FIS2011-22644) and TerMic
(FIS2014-52486-R) from the Spanish Government.  JMH is supported by the Gordon and Betty Moore Foundation through Grant GBMF4343.

\bibliographystyle{unsrt}
\bibliography{bib.bib}

\end{document}